\documentstyle[amssymb,amstex,11pt,graphics,epsf]{article}

\begin{document}

\title{Third-generation muffin-tin orbitals}
\author{O. K. Andersen, T. Saha-Dasgupta$^{\ast },$\ S. Ezhov \\
Max-Planck Institut f\"{u}r Festk\"{o}rperforschung, D-70569, Stuttgart,
Germany\\
$^{\ast }$S.N. Bose Centre, Kolkata 98, India}
\date{\today }
\maketitle

\begin{abstract}
By the example of $sp^{3}$-bonded semiconductors, we illustrate what
3rd-generation muffin-tin orbitals (MTOs) are. We demonstrate that they can
be downfolded to smaller and smaller basis sets: $sp^{3}d^{10},\;sp^{3},$
and bond orbitals. For isolated bands, it is possible to generate Wannier
functions {\it a priori.} Also for bands, which overlap other bands,
Wannier-like MTOs can be generated {\it a priori. }Hence, MTOs have a unique
capability for providing chemical understanding.

\end{abstract}

\section{Introduction}

Muffin-tin orbitals (MTOs) have been used for a long time in {\it ab initio,
e.g. }density-functional (DF), calculations of the electronic structure of
condensed matter. Over the years, several MTO-based methods have been
developed. The ultimate aim is to find a generally applicable
electronic-structure method which is {\em intelligible, fast,} and {\em %
accurate.}

In order to be {\em intelligible,} an electronic-structure method must
employ a {\em minimal} and {\em flexible} basis of {\em short-ranged}
orbitals. As an example, the method should be able to describe the valence
band and the lower part of the conduction band in $sp^{3}$-bonded materials
using merely four short-ranged $s$- and $p$-orbitals per atom and, for
insulating phases, using merely occupied orbitals such as bond orbitals.
Another example is materials with strong electronic correlations. For such
materials, one must first construct a small, but realistic Hilbert space of
many-electron wave functions, and this requires an accurate and flexible
single-particle basis of atom-centered short-ranged orbitals. A small basis
of short-ranged orbitals is a prerequisite for a method to be intelligible
and {\em fast,} but it may be a hindrance for its {\em accuracy,}\ because
the orbitals of a smaller basis tend to be more complicated than those of a
larger basis.
 
Most other density-functional methods, such as plane-wave pseudopotential,
LAPW, PAW, and LCAO methods, aim at {\em simulation,} and are therefore
primarily {\em accurate }and {\em robust.} But they are neither fast nor
intelligible in the above-mentioned sense, because they employ basis sets
with of order hundred functions per atom. With such methods, {\em %
understanding} can therefore only be attempted {\em after} the calculation,
by means of projections onto {\it e.g.} Wannier functions in case of
insulators,{\it \ }charge densities, electron-localization functions (ELFs),
partial waves, a.s.o..

The so-called 3rd-generation MTO method (Andersen {\it et al }1994, Andersen 
{\it et al} 1998, Andersen {\it et al} 2000, Tank and Arcangeli 2000,
Andersen and Saha-Dasgupta 2000) should come close to what we have been
aiming for. In the present paper we shall explain what 3rd-generation MTOs
are and what they achieve. Emphasis will be on the so-called {\em downfolding%
} and {\em energy-mesh} features which enable MTO bases to be small,
flexible, and accurate, as we shall demonstrate by exposing them to the
above-mentioned $sp^{3}$-test. From the result, the idea emerges, that for 
{\em band insulators,} an MTO basis can be designed {\it a priori} to span
the Hilbert space of the {\em occupied} states only. That is, there is one,
and only one, such MTO per electron. To get this count right, one may
associate each orbital with a nominal electron (or pair), and leave it to
the method to {\em shape} the orbitals in such a way that the basis set
becomes complete for the occupied states. This can be done because MTOs are 
{\em selective in energy, }in the sense that the MTOs of order N (NMTOs) are
shaped in such a way that the NMTO basis set solves Schr\"{o}dinger's
equation {\em exactly} for N+1 single-particle energies, which in the
present case must be chosen in such a way that they span the valence band.
This ability to generate Wannier functions directly in real space, should be
useful for {\em ab initio} molecular-dynamics simulations. NMTOs may also
prove useful for designing many-electron wave functions, which describe
correlated electron systems in a realistic way. The description of spin and
orbital ordering is a trivial example. Also for the conduction bands of {\em %
metals,} Wannier-like, low-energy MTOs can be designed {\it a priori.} This
has been demonstrated in several cases (M\"{u}ller {\it et al }1998, Sarma
and Saha-Dasgupta 2000, Korotin 2000, Valenti {\it et al} 2001, Dasgupta 
{\it et al} 2002), most recently for the hole-doped cuprate high-temperature
superconductors, where the material-dependent trend of the hopping integrals
and their correlation with the maximum $T_{c}$ was discovered (Pavarini {\it %
et al} 2001, Dasgupta {\it et al}).

For a description of how we expand the charge density locally in such a way
that Poisson's equation can be solved and the total energy and forces can be
evaluated fast and accurately, we refer to previous (Andersen {\it et al}
2000, Tank and Arcangeli 2000) and coming publications (Arcangeli and
Andersen, Savrasov and Andersen). This part of the 3rd-generation computer
code is still under construction.

\section{Screened spherical waves, kinked partial waves, and muffin-tin
orbitals}

The 3rd-generation MTO formalism is the multiple-scattering -- or KKR
(Korringa 1947, Kohn and Rostoker 1954)-- formalism for finding the
solutions, $\Psi _{i}\left( {\bf r}\right) ,$ of Schr\"{o}dinger's equation
for an electron in a muffin-tin potential, $V\left( {\bf r}\right)
=\sum_{R}v_{R}\left( r_{R}\right) ,$ with the following three extensions:

1.: The KKR formalism is proved to hold, not only for superpositions of
spherically-symmetric, non-overlapping potential wells, $v_{R}\left(
r_{R}\right) ,$ but also to leading order in the potential-{\em overlap}
(Andersen {\it et al} 1992). Here, and in the following, $r_{R}\equiv \left| 
{\bf r-R}\right| ,$ and ${\bf R}$ are the sites which we label by $R.$ The
potential, $v_{R}\left( r\right) ,$ is taken to vanish outside a radius, $%
s_{R},$ which should not exceed 1.6 times the radius of touching spheres. 
{\it i.e.:} $s_{R}+s_{R^{\prime }}\lesssim 1.6\left| {\bf R-R}^{\prime
}\right| $ for any pair of sites, $R$ and $R^{\prime }.$

2.: Exact {\em screening} transformations of the spherical waves, $%
n_{l}\left( \kappa r_{R}\right) Y_{L}\left( \hat{r}_{R}\right) ,$ are
introduced in order to reduce the spatial range and the energy dependence $%
\left( \kappa ^{2}\equiv \varepsilon \right) $ of the wave-equation
solutions, $\psi _{RL}\left( \varepsilon ,{\bf r}\right) $ (Andersen and
Jepsen 1984, Andersen {\it et al} 1992, Zeller {\it et al} 1995). Here, and
in the following, $L\equiv lm$ labels the spherical- (or cubic-) harmonic's
character.

3.: Energy-{\em in}dependent MTO basis sets are derived which span the
solutions $\Psi _{i}\left( {\bf r}\right) $ with energies $\varepsilon _{i}$
of Schr\"{o}dinger's equation to within errors proportional to $\left(
\varepsilon _{i}-\epsilon _{0}\right) \left( \varepsilon _{i}-\epsilon
_{1}\right) ..\left( \varepsilon _{i}-\epsilon _{N}\right) ,$ where $%
\epsilon _{0},\epsilon _{1},...,\epsilon _{N}$ is a chosen {\em energy mesh }%
with N+1 points (Andersen {\it et al} 2000, Andersen and Saha-Dasgupta
2000). Such an energy-independent set of Nth-order MTOs is called an NMTO
set. By virtue of the variational principle, the errors of the energies $%
\varepsilon _{i}$ will be proportional to $\left( \varepsilon _{i}-\epsilon
_{0}\right) ^{2}\left( \varepsilon _{i}-\epsilon _{1}\right) ^{2}..\left(
\varepsilon _{i}-\epsilon _{N}\right) ^{2}.$
 
\begin{figure}[b!]
\unitlength1cm
\begin{minipage}[ht]{15.4cm}
\centerline{
\rotatebox{0}{\resizebox{5.0in}{!}{\includegraphics{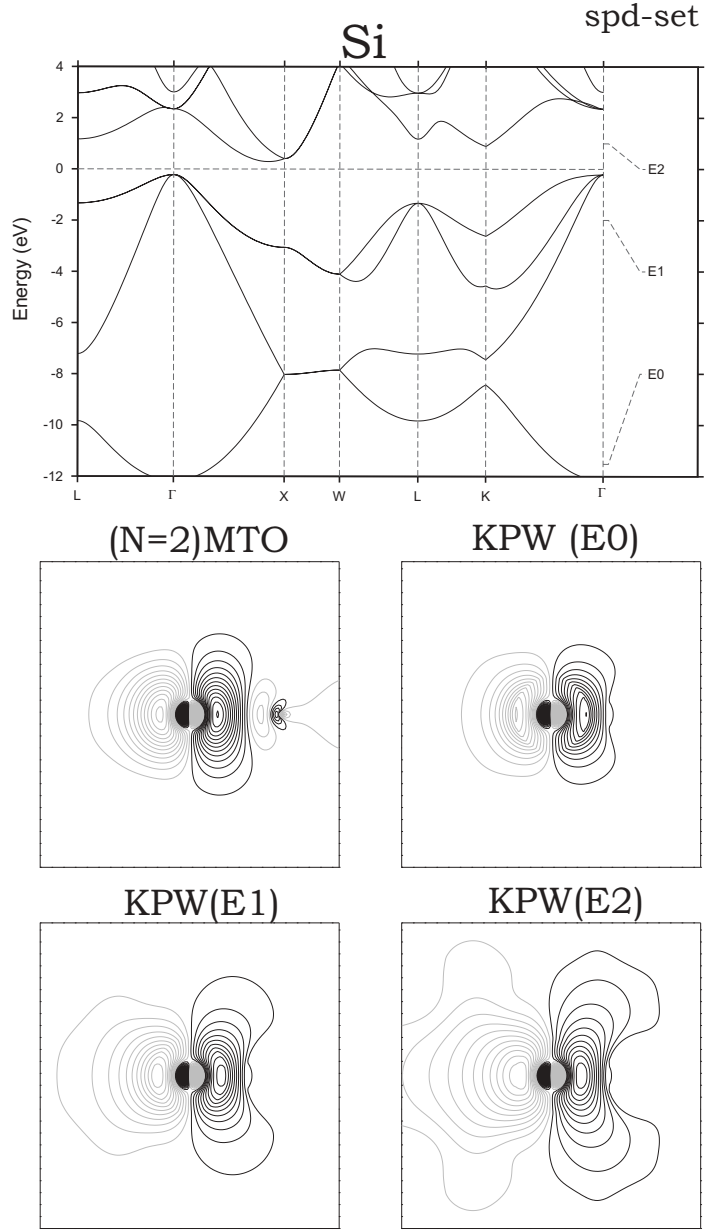}}}
}
\caption{
Band structure of Si calculated with the Si $spd$-QMTO basis set
corresponding to the energy mesh shown on the right-hand side (solid lines).
The contour plots show the Si $p$ orbital pointing in the [111]-direction
between two nearest neighbors in the (2\={1}\={1})-plane. Shown are the
kinked partial waves (KPWs) at the three energies and the QMTO. The KPWs are
normalized to 1, times a cubic harmonics, at the central hard sphere. The
contours are the same in all plots.
}
\end{minipage}
\hfill              
\end{figure}

At the top of figure 1 we show the LDA energy bands $\varepsilon _{i}\left( 
{\bf k}\right) $ of Si in the diamond structure, calculated with the basis
set of Si-centered $s$-, $p$-, and $d$-MTOs, {\it i.e. }with 9
orbitals/atom, for the 3-point energy mesh $\epsilon _{0},\,\epsilon
_{1},\,\epsilon _{2}$ indicated on the right-hand side. These bands have
meV-accuracy for the MT-potential, which in the present case was the
standard all-electron DF-LDA atomic-spheres potential. Since three energy
points were used, the MTOs are of order $N$=2, that is, they are quadratic
MTOs, so-called QMTOs.

The QMTO, $\chi _{p111}^{\left( 2\right) }\left( {\bf r}\right) ,$ pointing
along [111] from one Si to its nearest neighbor, is shown in the
(2\={1}\={1})-plane by the first contour plot. This orbital is localized and
smooth, with a few ''orthogonality wiggles'' at the nearest neighbor. The
remaining three contour plots show major constituents of this $p_{111}$%
-QMTO: The $p_{111}$-kinked partial wave (KPW), $\phi _{p111}\left( \epsilon
,{\bf r}\right) ,$ at the central site and at the three energies $\epsilon
_{0},\,\epsilon _{1},$ and $\epsilon _{2}.$ In figure 2 we show this $%
p_{111} $-QMTO, together with the $p_{111}$-KPW at the three energies, along
the line connecting the two nearest neighbors and proceeding into the
back-bond.
 
\begin{figure}[b!]
\unitlength1cm
\begin{minipage}[ht]{15.4cm}
\centerline{
\rotatebox{90}{\resizebox{5.0in}{!}{\includegraphics{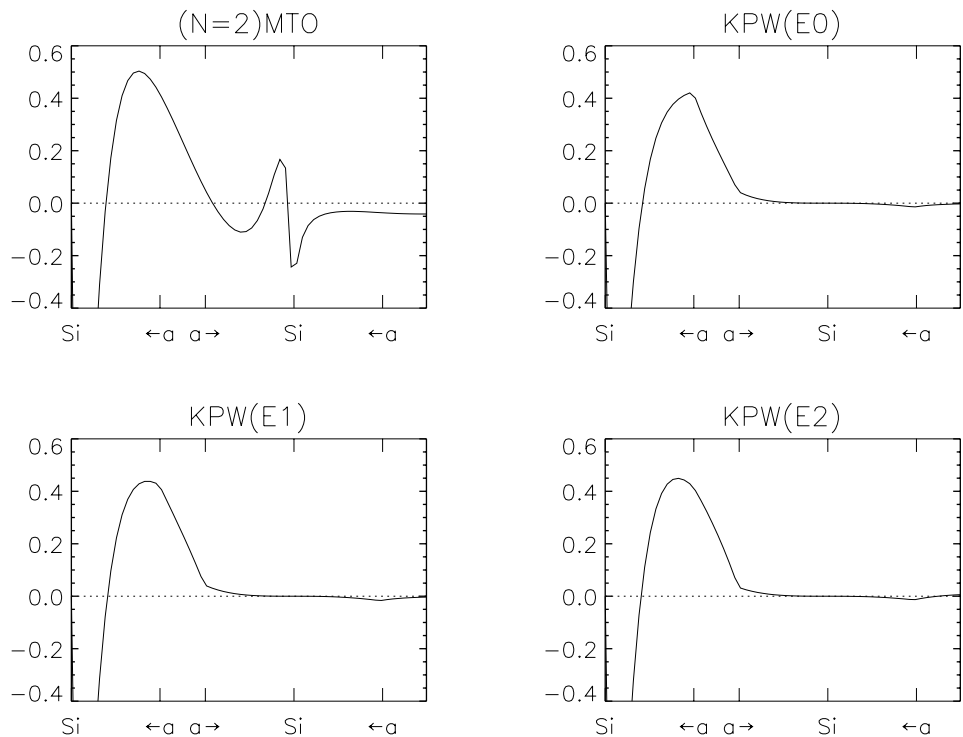}}}
}
\caption{
Si $p_{111}$-KPW for the Si $spd$-set plotted along the [111]-line
connecting nearest neighbors. The $a$'s indicate the hard spheres at the
central and the nearest-neighbor sites. The generating MT-potential had no
repulsive potential wells at interstitial sites (E), but only large
Si-centered wells with $s=1.7a.$
}
\end{minipage}
\hfill              
\end{figure}

In general, the members (labelled by $R^{\prime }L^{\prime })$ of the NMTO
basis set for the energy mesh $\epsilon _{0},...,\epsilon _{N}$ are
superpositions, 
\begin{equation}
\chi _{R^{\prime }L^{\prime }}^{\left( N\right) }\left( {\bf r}\right)
=\sum_{n=0}^{N}\sum_{RL\in A}\phi _{RL}\left( \epsilon _{n},{\bf r}\right)
\,L_{nRL,R^{\prime }L^{\prime }}^{\left( N\right) }\,,  \label{Lagrange}
\end{equation}%
of the kinked partial waves, $\phi _{RL}\left( \varepsilon ,{\bf r}\right) ,$
at the $N+1$ points (labelled by $n)$ of the energy mesh. In the present
case, the $L$-summation is over the nine $s$-, $p$-, and $d$-KPWs, and the $%
R $-summation is over all Si sites. Due to the localized nature of the KPWs
illustrated in the figures, the latter summation is limited to the
neighbors. The $RL$-values for which we have MTOs in the basis set, we label 
{\em active} $\left( A\right) ,$ or low. Expression (\ref{Lagrange}) is the
energy-quantized form of Lagrange interpolation,%
\[
\chi ^{\left( N\right) }\left( \varepsilon \right) \approx
\sum_{n=0}^{N}\phi \left( \epsilon _{n}\right) l_{n}^{\left( N\right)
}\left( \varepsilon \right) ,\quad l_{n}^{\left( N\right) }\left(
\varepsilon \right) \equiv \prod_{m=0,\neq n}^{N}\frac{\varepsilon -\epsilon
_{m}}{\epsilon _{n}-\epsilon _{m}}, 
\]%
of a function of energy, $\phi \left( \varepsilon \right) ,$ by an $N$%
th-degree polynomial, $\chi ^{\left( N\right) }\left( \varepsilon \right) :$
The $N$th-degree polynomial, $l_{n}^{\left( N\right) }\left( \varepsilon
\right) ,$ is substituted by a matrix with elements, $L_{nRL,R^{\prime
}L^{\prime }}^{\left( N\right) }\,,$ the function of energy, $\phi \left(
\varepsilon \right) ,$ by a Hilbert space with axes, $\phi _{RL}\left(
\varepsilon ,{\bf r}\right) ,$ and the interpolating polynomial, $\chi
^{\left( N\right) }\left( \varepsilon \right) ,$ by a Hilbert space with
axes, $\chi _{R^{\prime }L^{\prime }}^{\left( N\right) }\left( {\bf r}%
\right) .$

As illustrated in figures 1 and 2, a kinked partial wave is basically a
partial wave with a tail joined continuously to it with a {\em kink} at a
central, so-called hard sphere of radius $a_{R}$. This kink is seen most
clearly for the lowest energy, $\epsilon _{0}.$ As usual, the partial wave
is $\varphi _{Rl}\left( \varepsilon ,r_{R}\right) Y_{L}\left( {\bf \hat{r}}%
_{R}\right) ,$ where the function of energy is the regular solution of the
radial Schr\"{o}dinger equation, 
\begin{equation}
-\left[ r\varphi _{Rl}\left( \varepsilon ,r\right) \right] ^{\prime \prime }=%
\left[ \varepsilon -v_{R}\left( r\right) -l\left( l+1\right) /r^{2}\right]
r\varphi _{Rl}\left( \varepsilon ,r\right) ,  \label{radSchr}
\end{equation}%
for the potential-well $v_{R}\left( r\right) .$ The tail of the kinked
partial wave is a so-called {\em screened spherical wave}, $\psi _{RL}\left(
\varepsilon ,{\bf r}\right) ,$ which is essentially the solution with energy 
$\varepsilon $ of the wave equation in the interstitial between the hard
spheres, $-\Delta \psi \left( \varepsilon ,{\bf r}\right) =\varepsilon \psi
\left( \varepsilon ,{\bf r}\right) ,$ with the boundary condition that,
independent of the energy, $\psi _{RL}\left( \varepsilon ,{\bf r}\right) $
go to $Y_{L}\left( {\bf \hat{r}}_{R}\right) $ at the central hard sphere,
and to {\em zero} (with a kink) at all other hard spheres. It is this latter 
{\em confinement,} easily recognized in the plots, particularly at the
highest energy $\epsilon _{2},$ which makes the screened spherical waves,
the KPWs, and the MTOs localized when the energy is not too high. At the
same time, it makes the KPW have pure $L$-character merely at its central
sphere, because outside, it is influenced by the hard spheres centered at
the neighbors. The default value of the hard-sphere radii, $a_{R},$ is 90
per cent of the appropriate covalent, atomic, or ionic radius. The kinked
partial wave thus has a kink, not only at its own, but also at the
neighboring hard spheres, inside which it essentially vanishes.
'Essentially' because the above-mentioned boundary condition only applies to
the active components of the spherical-harmonics expansions of the screened
spherical wave on the hard spheres. For the remaining components, in the
present case the Si $f$- and higher components, as well as all components on
empty (E) spheres, the screened spherical wave equals the corresponding
partial-wave solution of Schr\"{o}dinger's equation throughout the
MT-sphere. The small bump seen in figure 1 in the lowest KPW contour along
the [111]-direction is mainly caused by the $f$-character on the nearest
neighbor, and so is the finite amplitude seen in figure 2 inside the nearest
hard sphere.

\section{Computational steps}

The radial Schr\"{o}dinger (Dirac) equations (\ref{radSchr}) are integrated
numerically from $r=0$ to $s_{R}.$ This yields the radial functions, $%
\varphi _{Rl}\left( \varepsilon ,r\right) ,$ and their phase shifts, $\eta
_{Rl}\left( \varepsilon \right) ,$ each of which are obtained by matching
the logarithmic derivative of $\varphi _{Rl}\left( \varepsilon ,r\right) $
at $r=s_{R}$ to that of%
\begin{equation}
\varphi _{Rl}^{o}\left( \varepsilon ,r\right) \propto j_{l}\left( \kappa
r\right) -\tan \eta _{Rl}\left( \varepsilon \right) n_{l}\left( \kappa
r\right) .  \label{phizero}
\end{equation}%
The radial integration must be performed for each potential well and for
each $l,$ increasing until all further phase shifts vanish due to dominance
of the centrifugal term in (\ref{radSchr}).

The screened spherical waves are specified by a Hermitian structure matrix,
whose element $B_{R^{\prime }L^{\prime },RL}\left( \varepsilon \right) $ is
essentially the radial logarithmic derivative of the $L^{\prime }$-component
in the spherical-harmonics expansion at the hard sphere at site $R^{\prime }$
of the screened spherical wave $\psi _{RL}\left( \varepsilon ,{\bf r}\right)
.$ What is known analytically, is the element%
\[
B_{R^{\prime }L^{\prime },RL}^{0}\left( \varepsilon \right) \equiv
\sum_{l"}4\pi i^{-l+l^{\prime }-l^{\prime \prime }}C_{LL^{\prime }l^{\prime
\prime }}\kappa n_{l^{\prime \prime }}\left( \kappa \left| {\bf R-R}^{\prime
}\right| \right) Y_{L^{\prime \prime }}^{\ast }\left( \widehat{{\bf R-R}%
^{\prime }}\right) , 
\]%
of the {\em bare} KKR structure matrix, which specifies how the spherical
wave, $n_{l}\left( \kappa r_{R}\right) Y_{L}\left( \hat{r}_{R}\right) ,$ at
site $R$ is expanded around another site, $R^{\prime },$ in regular
spherical waves, $j_{l^{\prime }}\left( \kappa r_{R^{\prime }}\right)
Y_{L^{\prime }}\left( \hat{r}_{R^{\prime }}\right) .$ Here $\kappa
^{2}\equiv \varepsilon ,$ and the on-site terms of the bare structure matrix
are defined to vanish. Screening of the structure matrix,%
\[
\left[ B\left( \varepsilon \right) ^{-1}\right] _{RL,R^{\prime }L^{\prime
}}\equiv \left[ B^{0}\left( \varepsilon \right) ^{-1}\right] _{RL,R^{\prime
}L^{\prime }}+\kappa ^{-1}\tan \alpha _{RL}\left( \varepsilon \right) \delta
_{RR^{\prime }}\delta _{LL^{\prime }}, 
\]%
requires inversion of the matrix $B_{RL,R^{\prime }L^{\prime }}^{0}\left(
\varepsilon \right) +\kappa \cot \alpha _{RL}\left( \varepsilon \right)
\delta _{RR^{\prime }}\delta _{LL^{\prime }}.$ This can be done by fixing $R$
and limiting $R^{\prime }$ to the 10-50 nearest sites. $\alpha _{RL}\left(
\varepsilon \right) $ are the hard-sphere phase shifts for the active
channels,%
\[
\tan \alpha _{RL}\left( \varepsilon \right) \equiv j_{l}\left( \kappa
a_{R}\right) /n_{l}\left( \kappa a_{R}\right) , 
\]%
and, for the remaining channels, $\alpha _{RL}\left( \varepsilon \right) $
are the proper phase shifts, $\eta _{RL}\left( \varepsilon \right) .$ The
latter channels, which will not have KPWs and MTOs associated with them, are
said to be {\em downfolded.} With appropriate division into active and
downfolded channels, the screened structure matrix will have short spatial
range and no poles in the energy-range of the occupied states.

A kinked partial wave is defined as:%
\begin{equation}
\phi _{RL}\left( \varepsilon ,{\bf r}\right) =\left[ \varphi _{Rl}\left(
\varepsilon ,r_{R}\right) -\varphi _{Rl}^{o}\left( \varepsilon ,r_{R}\right) %
\right] Y_{L}\left( \hat{r}_{R}\right) +\psi _{RL}\left( \varepsilon ,{\bf r}%
\right) ,  \label{KPW}
\end{equation}%
where $\varphi \left( \varepsilon ,r\right) $ is the radial solution for the
central well from 0 to $s,$ and $\varphi ^{o}\left( \varepsilon ,r\right) $\
is the phase-shifted wave (\ref{phizero}) proceeding smoothly {\em inwards}
from $s$ to the central $a$-sphere, where it is matched with a kink to the
screened spherical wave $\psi \left( \varepsilon ,{\bf r}\right) .$ The
kinks of the KPW set are then given by the kink matrix,%
\begin{equation}
K_{RL,R^{\prime }L^{\prime }}\left( \varepsilon \right) \equiv \frac{%
B_{RL,R^{\prime }L^{\prime }}\left( \varepsilon \right) +\kappa \cot \eta
_{RL}^{\alpha }\left( \varepsilon \right) \delta _{RR^{\prime }}\delta
_{LL^{\prime }}}{-\varepsilon n_{l}\left( \kappa a_{R}\right) n_{l^{\prime
}}\left( \kappa a_{R^{\prime }}\right) },  \label{K}
\end{equation}%
where $\eta _{RL}^{\alpha }\left( \varepsilon \right) $ is the phase shift
with respect to the hard-sphere medium,%
\[
\tan \eta _{RL}^{\alpha }\left( \varepsilon \right) \equiv \tan \eta
_{RL}\left( \varepsilon \right) -\tan \alpha _{RL}\left( \varepsilon \right)
. 
\]%
The rows and columns of the kink matrix run merely over active channels. In
the fomalism above, we have for simplicity used the notation of scattering
theory, which is analytical for $\varepsilon >0,$ and for $\varepsilon <0.$ 
{\em Screened} scattering theory with the normalization (\ref{K}) is however
analytical in a region of interest {\em around} $\varepsilon =0.$

Finally, the Lagrange matrix which gives the MTO set (\ref{Lagrange}) in
terms of the KPW set (\ref{KPW}), is given solely in terms of the values of
the Green matrix, $G\left( \varepsilon \right) \equiv K\left( \varepsilon
\right) ^{-1}$, on the energy mesh $\varepsilon =\epsilon _{0},\epsilon
_{1},...,\epsilon _{N}.$ The Hamiltonian and overlap matrices in the MTO
representation are given in terms of the same values of the Green matrix,
plus the values of its first energy derivative, $\dot{G}\left( \varepsilon
\right) .$

\section{Further downfoldings}

In the valence and lowest conduction bands of Si, there are only $s$- and $p$%
-, but no $d$-electrons. To describe these bands, we should therefore be
able to use a basis with only Si $s$- and $p$-MTOs, that is, with only 4
orbitals per atom. We thus let the Si $s$- and $p$-partial waves remain
active, while the Si $d$-waves are now included among the passive ones, {\it %
i.e.} those 'folded down' into the tails of the screened-spherical waves in (%
\ref{KPW}). The results for the bands and the $p_{111}$-QMTO are shown in
figures 3 and 4. These bands are indistinguishable from those obtained with
the Si $spd$-set, on the scale of the figure, although between the energies
of the mesh, the former bands do lie slightly above the latter. However, by
making the mesh denser (increasing $N),$ the accuracy can be increased
arbitrarily. The KPW of the $sp$-set is seen to have $d$-character on the
nearest Si neighbor, and the QMTO and the KPW, particularly the one at the
highest energy, are seen to be somewhat less localized than those for the $%
spd$-set.
 
\begin{figure}[b!]
\unitlength1cm
\begin{minipage}[ht]{15.4cm}
\centerline{
\rotatebox{0}{\resizebox{5.0in}{!}{\includegraphics{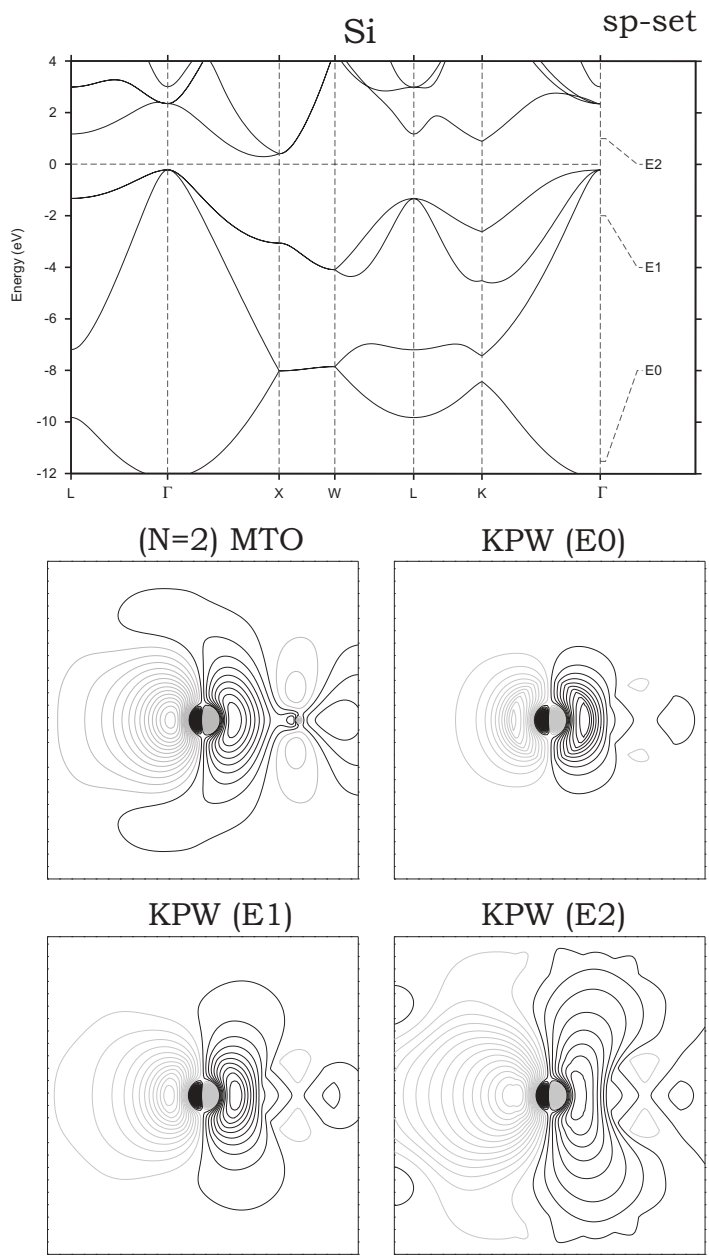}}}
}
\caption{
Same as figure 1, but for the Si $sp$-set.
}
\end{minipage}
\hfill              
\end{figure}
 
\begin{figure}[b!]
\unitlength1cm
\begin{minipage}[ht]{15.4cm}
\centerline{
\rotatebox{90}{\resizebox{5.0in}{!}{\includegraphics{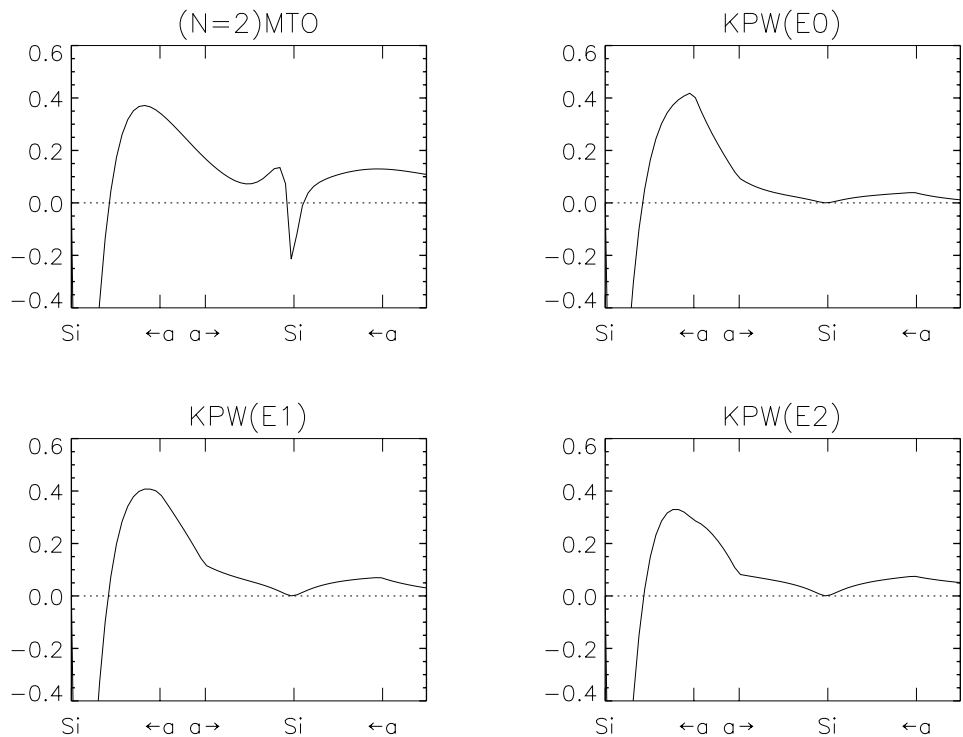}}}
}
\caption{
Same as figure 2, but for the Si $sp$-set.
}
\end{minipage}
\hfill              
\end{figure}

It is even possible to construct an arbitrarily accurate MTO-basis which
spans merely the {\em occupied} orbitals, that is, which spans the valence
band with a basis of merely 2 orbitals per atom. For tetrahedrally
coordinated covalent semiconductors like Si, it is customary to take the
valence-band orbitals as the bond-orbitals, which are the bonding linear
combinations of directed $sp^{3}$-hybrids of orthonormal orbitals. It is,
however, far simpler and more general, {\it e.g.} not limited to elemental
semiconductors and tetrahedral structures, to take the valence-band orbitals
as the $s$- and $p$-MTOs on {\em every second} Si atom, all partial waves on
the nearest neighbors being downfolded. This corresponds to a Si$^{4+}$Si$%
^{4-}$ {\em ionic} picture. This QMTO-set turns out to describe merely the
valence band, and to do so surprisingly well considering the fact that the
two silicons are treated differently, so that the degeneracy along the
XW-line is, in fact, slightly broken. The error between the energy points is
proportional to $\left[ \varepsilon _{i}\left( {\bf k}\right) -\epsilon _{0}%
\right] \left[ \varepsilon _{i}\left( {\bf k}\right) -\epsilon _{1}\right] %
\left[ \varepsilon _{i}\left( {\bf k}\right) -\epsilon _{2}\right] ,$
exactly as for the basis with 4 orbitals per atom shown in figure 3, because
we use QMTOs in both cases, but the prefactors are larger for the smaller
basis: As the number of active channels decreases, the KPWs attain longer
range and stronger energy dependence. However, by making the energy mesh
finer, the errors of the MTO set can be made arbitrarily small. In the
bottom line of figure 5, we show the result of such a valence-band-only
calculation with $N=3$ for Ge, together with the $p_{111}$ cubic MTO (CMTO)
centered on the Ge atom to the right. The accuracy of the valence band is
superb, and the MTO is seen to spill over onto the nearest-neighbor atom(s)
which were chosen not to have orbitals associated with them.
 
\begin{figure}[b!]
\unitlength1cm
\begin{minipage}[ht]{15.4cm}
\centerline{
\rotatebox{0}{\resizebox{5.0in}{!}{\includegraphics{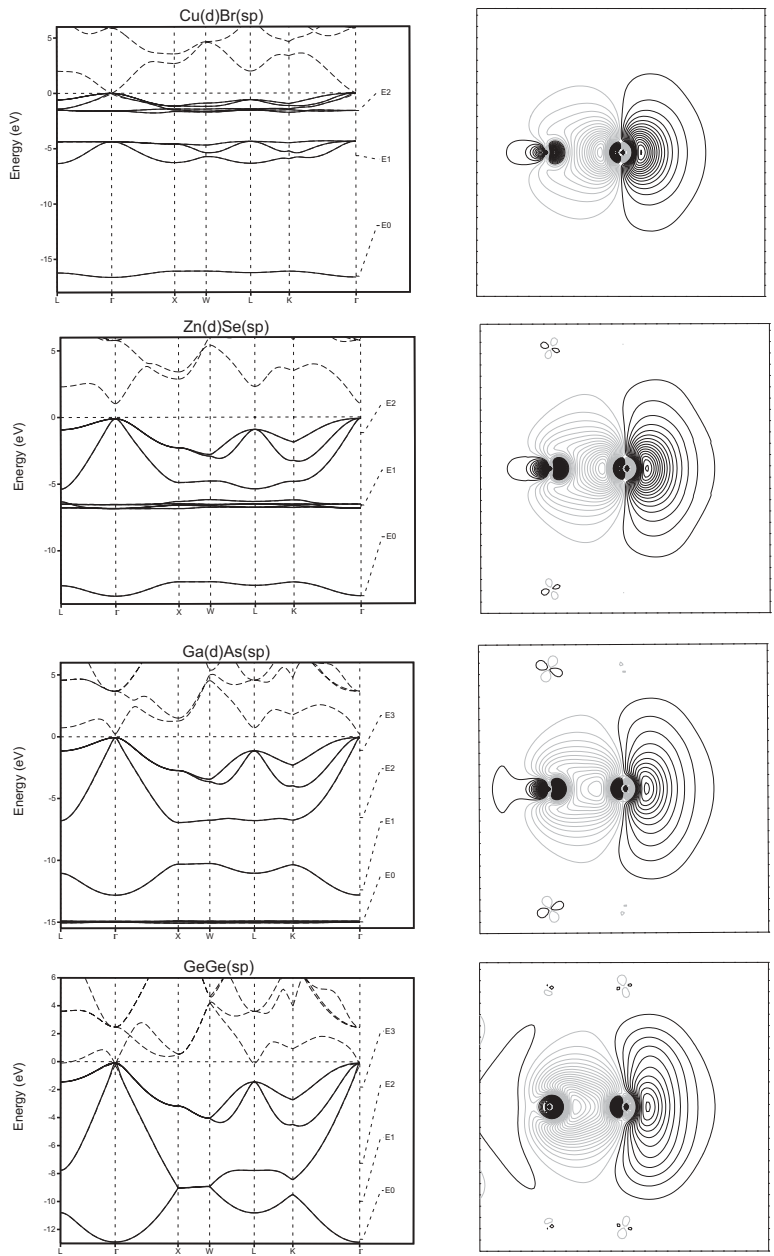}}}
}
\caption{
Valence bands (solid lines) of the series CuBr -- Ge, calculated
with the ionic basis sets where the $sp$-MTOs are on the anion and, except
for Ge, the $d$-MTOs are on the cation. The exact bands are given by dashed
lines. The contour plots show the $p_{111}$-MTO on the anion (the atom to
the right). The ionicity decreases and the covalency increases from the top
to the bottom.
}
\end{minipage}
\hfill              
\end{figure}

Since this basis is {\em complete} for the {\em occupied} states, we may
compute the density-functional ground-state properties in {\em real space}
by taking traces, provided that we first L\"{o}wdin orthonormalize the basis
in real space. The sum of the one-electron energies is then computed as the
trace of the Hamiltonian, {\it i.e.} as the sum of the energies of the
orthonormal orbitals, and the charge density is computed as the sum of the
squares of these orbitals. This is a method where the amount of computation
increases merely linearly with the size of the system, a so-called {\em %
order-}${\cal N}${\em \ method. }Here, ${\cal N}$ refers to the number of
atoms in the system and not to the order $N$ of the MTOs. This NMTO method,
which generates the complete basis for the occupied states {\it a priori, }%
should be superior to current order-${\cal N}$ methods, which either use
inaccurate empirical tight-binding models or project onto the occupied
states during the course of a large-basis-set calculation.
 
\begin{figure}[b!]
\unitlength1cm
\begin{minipage}[ht]{15.4cm}
\centerline{
\rotatebox{0}{\resizebox{5.0in}{!}{\includegraphics{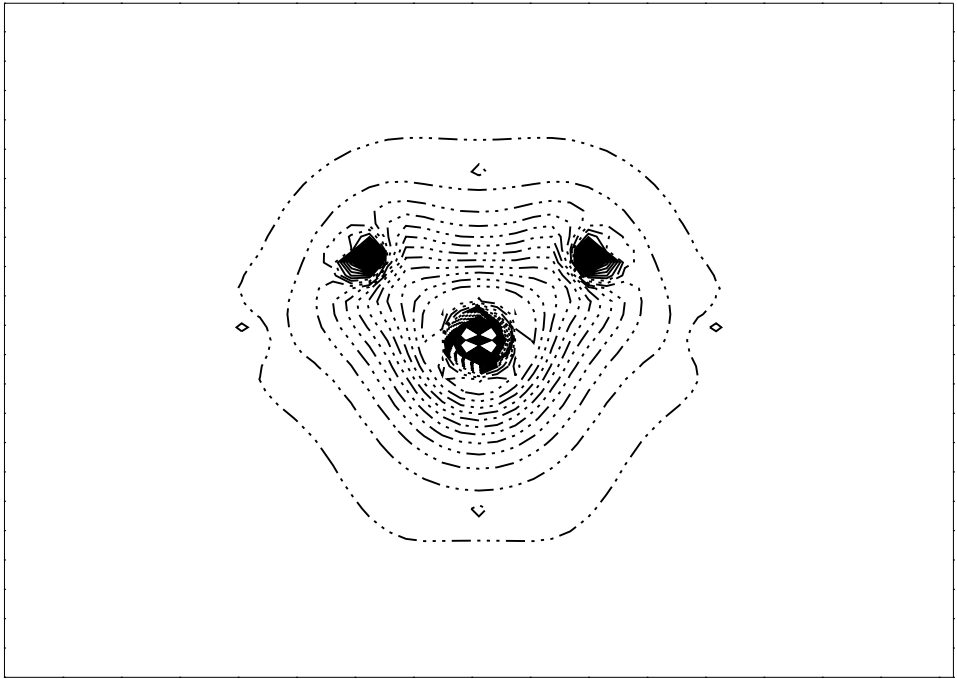}}}
}
\caption{
Contour plot in the $\left( 1\bar{1}0\right) $-plane of the
orthogonalized $s$-QMTO on the central Si atom. All partial waves on the
nearest neighbors (and their lattice translations) were downfolded.
}
\end{minipage}
\hfill              
\end{figure}
 
\begin{figure}[b!]
\unitlength1cm
\begin{minipage}[ht]{15.4cm}
\centerline{
\rotatebox{0}{\resizebox{5.0in}{!}{\includegraphics{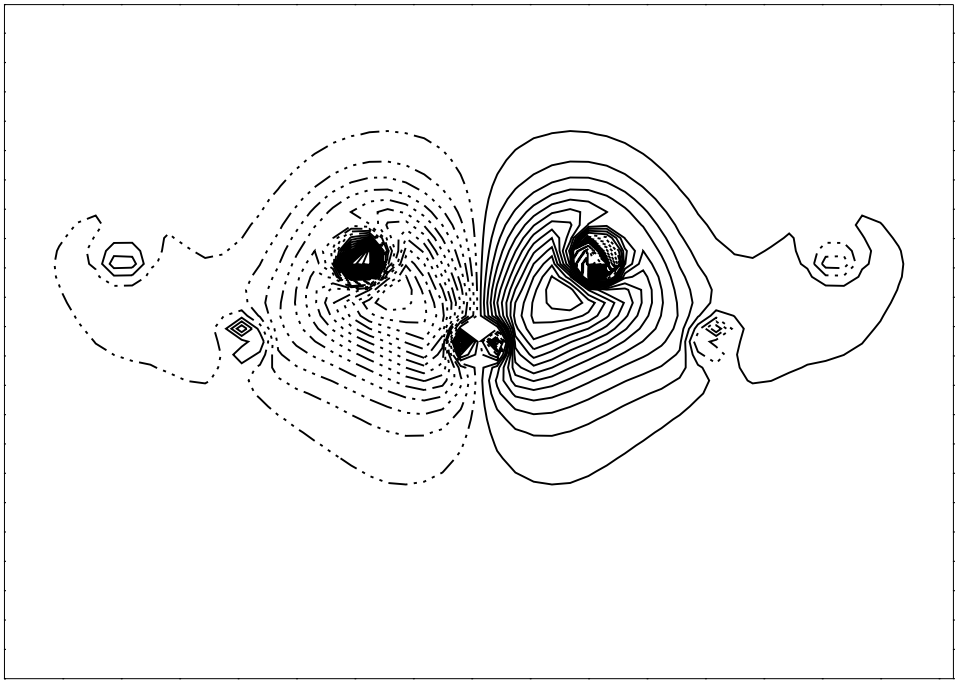}}}
}
\caption{
Same as figure 6, but for the $p_{x}$-QMTO.
}
\end{minipage}
\hfill              
\end{figure}

In order to demonstrate in further detail that our method works, we consider
Si in the diamond structure for which the valence-band Wannier functions can
be taken as {\em bond orbitals.} First, we orthonormalize our
symmetry-breaking Si$^{4+}$ Si $s,p_{x},p_{y},p_{z}$ QMTO set. The resulting
Si $s$ and Si $p_{x}$ orbitals are shown in the (110)-plane in figures 6 and
7. These orthogonalized orbitals are seen to remain fairly localized. Then,
we transform to the four congruent $sp^{3}$-hybrids centered nominally on
every second Si atom. As figure 8 shows, such an $sp^{3}$-hybrid is, in
fact, the bond orbital. Hence, folding all partial waves of the atoms chosen
not to carry orbitals, into the tail of the $sp^{3}$-directed orbital on one
of the other Si atoms, has made that orbital look like a bond orbital. The
reason why the figure does not show exact symmetry between the two sites is
caused by our use of an energy mesh with only 3 points in the valence band.
Making the energy mesh finer will generate the exact symmetry.
 
\begin{figure}[b!]
\unitlength1cm
\begin{minipage}[ht]{15.4cm}
\centerline{
\rotatebox{0}{\resizebox{5.0in}{!}{\includegraphics{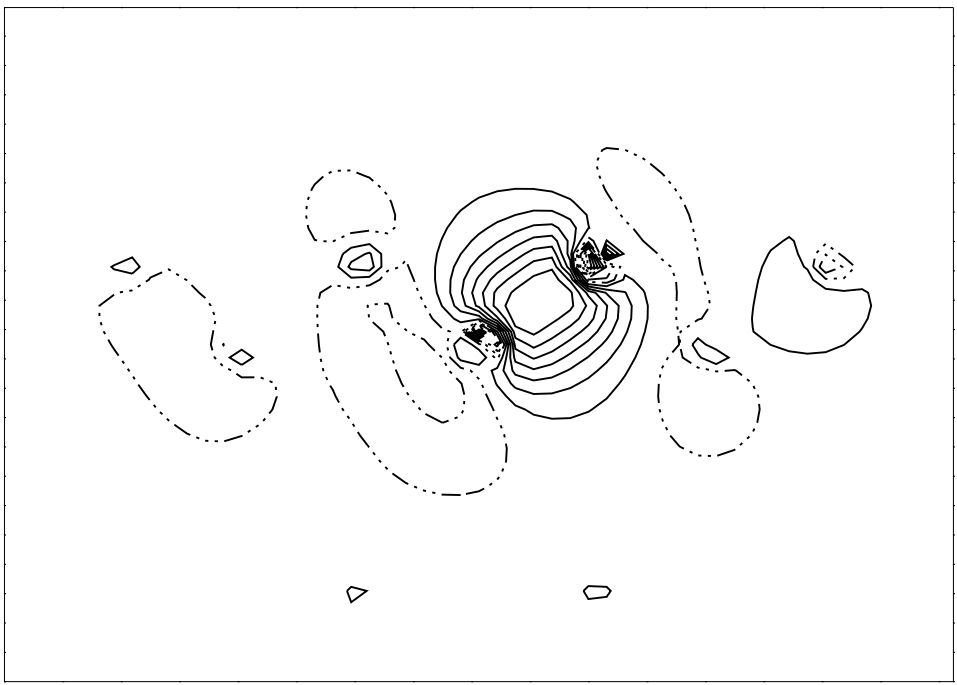}}}
}
\caption{
Same as figure 6, but for the $sp^{3}$-hybrid. For increasing $N,$
this orbital converges to the bond orbital.
}
\end{minipage}
\hfill              
\end{figure}

Now, Sn is metal because the bonding and antibonding bands overlap. This
should, however, not prevent our method from working for the occupied states
only, because NMTOs are {\em energy selective.} Since the orbitals will be
shaped in such a way that the basis set solves Schr\"{o}dinger's equation
exactly for the energies on the mesh, we may merely have to choose several
energy points in the region of band overlap {\em below} the Fermi level
--and, of course, {\em no} energy points above$.$ Remember that our ionic
prescription does not make use of the fact that the Wannier functions for
the valence and conductions bands are respectively bonding and antibonding.

Since the ionic Si Si$\left( sp\right) $ set gives the occupied states in
diamond-structured Si with arbitrary accuracy, the same procedure with the $%
sp$-orbitals placed exclusively on the anion, and the $d$-orbitals on the
cation, will of course work for any IV-IV, III-V, II-VI, and I-VII
semiconductor and insulator. CuBr, for instance, would be thought of as an
ionic compound Cu$^{+}$Br$^{-}$ with the closed-shell configuration Cu$%
3d^{10}$ Br$4s^{2}4p^{6},$ and the basis should therefore have the $d$-MTOs
on the Cu atoms and the $s$ and $p$-MTOs on the Br atoms. This is
illustrated in the upper line of figure 5. Being a single-site prescription,
this works for CuBr in {\em any} structure. What we have seen is thus, that
the MTO basis can be designed {\it a priori} to span the Hilbert space of
the occupied states only. That is, there is one, and only one, such MTO per
electron. To specify such a set, one would, in order to get the
electron-count right, put the orbitals where the electrons are though to be,
and leave it to the method to {\em shape} the tails of these orbitals in
such a way that the basis solves Schr\"{o}dinger's equation exactly for
occupied states of the given static mean-field ({\it e.g.} LDA) potential.
Such {\em ionic} MTO {\em basis sets,} which 'automatically' span the
occupied --and no further-- states of any {\em band insulator,} could make
density-functional molecular-dynamics calculations highly efficient for such
systems.

How could one imagine to treat a chemical reaction like: 2$H_{2}O\rightarrow
2H_{2}+O_{2}$ with MTO bases of occupied states only? For water, it is
natural to use the ionic description H$_{2}^{+}$O$^{--},$ according to which
the orbital configuration is O $p^{6},$ {\it i.e.} one would put the $p$
orbitals on oxygen and fold down all partial waves centered on the
hydrogens. In principle, one might stick to this configuration throughout
the reaction, because it keeps the electron-count right. However, the
oxygen-centered orbitals would eventually look strange and have long range,
because they would have to separate off pieces of wave functions sitting on
the hydrogens. Such oxygen orbitals might be more time-consuming to
generate. At some stage in the reaction, it might therefore be appropriate
to switch to configurations such as H$s\uparrow $ H$s\downarrow ,$ or H H$%
s^{2}$ for the hydrogen molecule; this is analogous to our treatment of the
occupied states in tetrahedrally coordinated Si. For O$_{2}$ with the {\em %
open-shell} molecular configuration $pp\sigma ^{2}$ $pp\pi ^{4}$ $pp\pi
\uparrow \uparrow ,$ we might use an 'ionic' configuration like: O $\left(
z\uparrow \,x\uparrow \downarrow \,y\uparrow \right) $ O $\left( z\downarrow
\,x\uparrow \,y\uparrow \downarrow \right) $ with $z$ referring to the {\em %
local} $z$-direction of the molecule.

This example immediately leads to a treatment of {\em open-}shell systems by
means of spin and possibly orbital polarizations. We have seen that the
3rd-generation MTO method offers the possibility of designing
single-electron bases of atom-centered localized orbitals, which span the
wave functions in a given energy region of a given mean-field potential.
These orbitals can even be {\em symmetry-breaking}, as in the case of
diamond, Si, Ge, and Sn, without the generating mean field having to be so.
These orbitals thus seem to have great potential in the design of {\em %
many-electron wave functions} which describe {\em correlated electron}
systems in a {\em realistic} way.

\section{Conclusion}

We have solved the long-standing problem of deriving energy-independent,
short-ranged orbitals from scattering theory (Hubbard 1967). The present
formalism contains exactly the right 'physics and chemistry,' we feel. This
should give the computational method great speed and accuracy, and make it a
vehicle for discovery and understanding.

\section{References}

Andersen O K and Jepsen O 1984 {\it Phys. Rev. Lett.} {\bf 53} 2571

Andersen O K, Postnikov A V, and Savrasov S Yu 1992 {\it Applications of
Multiple Scattering Theory to Materials Science} eds. W H Butler, P H
Dederichs, A Gonis and R L Weaver, Mat. Res. Soc. Symp. Proc. Vol. {\bf 253}
(Pittsburgh: Materials Research Society) 37-70

Andersen O K, Jepsen O, and Krier G 1994 {\it Lectures on Methods of
Electronic Structure Calculations} eds V. Kumar, O.K. Andersen, and A.
Mookerjee (Singapore: World Scientific Publishing Co) 63-124

Andersen O K, Arcangeli C, Tank R W, Saha-Dasgupta T, Krier G, Jepsen O, and
Dasgupta I 1998 {\it Tight-Binding Approach to Computational Materials
Science }eds L Colombo, A Gonis, and P Turchi, Mat. Res. Soc. Symp. Proc.
Vol. {\bf 491} (Pittsburgh: Materials Research Society) 3-34

Andersen O K, Saha-Dasgupta T, Tank R W, Arcangeli C, Jepsen O, and Krier G
2000 {\it Electronic Structure and Physical Properties of Solids. The Uses
of the LMTO Method }ed H Dreysse (New York: Springer Lecture Notes in
Physics) 3-84

Andersen O K and Saha-Dasgupta T 2000 {\it Phys. Rev.} B {\bf 62} R16 219

Arcangeli C and Andersen O K {\it to be published}

Dasgupta I {\it et al }2002 {\it Bull Mater Sci }this issue

Dasgupta I, Saha-Dasgupta T, Andersen O K, Pavarini E, and Jepsen O{\it \ to
be published}

Hubbard J 1967 {\it Proc. Phys. Soc. London} {\bf 92 }921, and references
therein

Kohn W and Rostoker J 1954 {\it Phys. Rev.} {\bf 94} 111

Korotin M A, Anisimov V I, Saha-Dasgupta T, Dasgupta I 2000 {\it Journal of
Physics: Condensed Matter} {\bf 12} 113-124

Korringa J 1947 {\it Physica} {\bf 13} 392

M\"{u}ller T F A, Anisimov V, Rice T M, Dasgupta I, Saha-Dasgupta T 1998 
{\it Phys. Rev.} B {\bf 57} R12655

Pavarini E, Dasgupta I, Saha-Dasgupta T, Jepsen O, and Andersen O K 2001 
{\it Phys. Rev. Lett.} {\bf 87} 047003

Sarma D D, Mahadevan P, Saha-Dasgupta T, Ray S, and Kumar A 2000 {\it Phys.
Rev. Lett.} {\bf 85} 2549

Savrasov D and Andersen O K {\it to be published}

Tank R W and Arcangeli C 2000 {\it phys. stat. sol.} (b) {\bf 217} 89-130

Valenti R, Saha-Dasgupta T, Alvarez J V, Pozgajcic K, and Gros C{\it \ }2001 
{\it Phys. Rev. Lett. }{\bf 86, }5381 (2001)

Zeller R, Dederichs P H, Ujfalussy B, Szunyogh L, and Weinberger P{\it \
1995 Phys. Rev.} B {\bf 52} 8807

\end{document}